\documentstyle[prl,aps,epsf,twocolumn]{revtex}
\epsfverbosetrue
\begin{document}
\draft
\twocolumn[\hsize\textwidth\columnwidth\hsize\csname  
@twocolumnfalse\endcsname
\author{Jiri Maly, Boldizs\'ar Jank\'o,  and K. Levin}
\address{The James Franck Institute, The University of Chicago, 5640
South Ellis Avenue, Chicago IL 60637 USA}
\date{May 1, 1998}
\title{ Pairing Correlations  and the Pseudo-Gap 
State: Application of the ``Pairing Approximation'' Theory
}
\maketitle
\hspace*{-0.25ex}\begin{abstract}
We investigate the pseudogap onset temperature $T^*$, the
superconducting transition temperature $T_c$ and the general nature of
the pseudogap phase using a diagrammatic BCS-Bose Einstein crossover
theory. This decoupling scheme is based on the ``pairing
approximation'' of Kadanoff and Martin, further extended by Patton
(KMP).  Our consideration of the KMP ``pairing approximation'' is
driven by the objective to obtain BCS like behavior at weak coupling,
(which does not necessarily follow for other diagrammatic
schemes). Two coupled equations, corresponding to those for the single
particle and pair propagators, must be solved numerically, along with
the number equation constraint.  The variation of small to large
coupling constant $g$ is explored, whereby the system is found to
cross over from BCS to Bose-Einstein behavior. Our numerical
calculations proceed in two stages: first we investigate the ``lowest
order theory'', which is appropriate at temperatures well above
$T_c$. We use this theory to determine where the Fermi liquid state
first breaks down. This breakdown, which occurs at $T^*$ and is
associated with intermediate values of the coupling, corresponds to a
splitting of the single peaked (Fermi liquid) electronic spectral
function into two peaks well separated by a gap, as might be expected
for the pseudogap phase.  Indeed, our calculations provide physical
insight into the pseudogap state which is characterized by the
presence of metastable pairs or ``resonances'', which occupy states
around the Fermi energy; in this way they effectively reduce the
single particle density of states. The superconducting instability
$T_c$ is evaluated in the second stage of our calculations. Here we
introduce ``mode coupling" effects, in which the long lived pairs are
affected by the single particle pseudogap states and vice versa. Our
$T_c$ equations, which turn out to be rather simple as a result of the
KMP scheme, reveal a rich structure as a function of $g$ in which the
pseudogap is found to compete with superconductivity. Our results are
compared with alternate theories in the literature.
\end{abstract}
\hspace*{-0.25ex}
\hfill {\tt cond-mat/9805018}
]

\narrowtext

\section{Introduction}
\label{Section_1}

Pseudogap phenomena were reported in the first generation of
underdoped high $T_c$ samples\cite{magnetic}, although they received
little attention at that time. These early experiments\cite{magnetic}
on underdoped cuprates demonstrated a suppression in the magnetic
susceptibility with decreasing temperature ($T < T^*$) which was
suggestive of some form of singlet pairing in the normal state.  These
first indications of the pseudogap state were strictly associated with
magnetic probes. Hence the phrase ``spin gap'' was used to describe
them and related frequency dependent magnetic
phenomena\cite{neutrons}. More recently, it has become clear that
similar effects\cite{charge} are present in the charge response as
well.  Today the spin gap is more broadly interpreted to correspond to
some form of excitation gap (i.e., pseudogap) in both charge and spin
channels. It can and has been debated as to whether the effects are
equally profound in each.

Among the most revealing data on the pseudogap state are angle
resolved photoemission experiments\cite{arpesanl,arpesstanford}
(ARPES) which demonstrate that below $T^*$ the pseudogap is present in
the energy dispersion of the normal state electrons and that,
moreover, it seems to have the same (d-wave) symmetry as the
superconducting state, and evolves more or less continuously through
$T_c$ into a superconducting excitation gap. Finally, these data show
that the pseudogap is attached to an underlying Luttinger volume Fermi
surface, although there are many aspects about the high temperature
regime ($T> T^*$), (where the pseudogap is essentially absent), which
do not correspond to a canonical Fermi liquid.

Theories of the cuprate pseudogap can be broadly divided into two
groups: those that associate this state with magnetic pairing of some
sort\cite{Lee,Anderson,Fukuyama,Pines,Schrieffer,Chubukov} and those
which view the pseudogap as deriving from some form of precursor
superconductivity
\cite{Emery,Randeria,SadeMelo,Micnas,Ranninger,Alexandrov,TDLee,Tchern,Uemura,Larkin,Janko,Maly}.
(In reality the division between these two classes is somewhat
blurred, since in many scenarios magnetic pairing is a natural
requisite for superconductivity).  The present paper belongs to this
second category. It is our contention that this second scenario
deserves careful attention in large part because mean field
descriptions of the transition (which neglect precursor or pairing
correlation effects) are not expected to be applicable to short
coherence length, quasi 2d superconductors such as the
cuprates. Moreover, ARPES data suggest a smooth evolution of the
pseudo- into the superconducting gap.  Finally, these systems
ultimately undergo a superconducting, not magnetic transition. At the
very least it is clear that the pseudogap state will never be fully
characterized unless precursor superconducting effects are properly
calibrated.

Within the precursor superconductivity school, there are, moreover,
two distinct classes. The first of these is based on the short
coherence length $\xi$ which is naturally associated with modest size
Cooper pairs and, therefore, a superconducting state intermediate
between the BCS (large pair) and Bose-Einstein (small pair)
descriptions.  The second class is based on the anomalously low plasma
frequency $\omega_p$ which suggests that phase fluctuations play an
important role. In this way the amplitude of the order parameter is
established at $T^*$, while phase coherence occurs at the much lower
temperature $T_c$. In reality, both $\xi$ and $\omega_p$ are small and
some initial work by our group\cite{MalyCoulomb} has treated these two
parameters on a more even footing.

While, there have been no decisive arguments in favor of one or the
other of the two precursor superconductivity schools, in this paper we
explore the ``BCS Bose-Einstein crossover'' scenario, in large part
because it lends itself to more detailed diagrammatic and therefore
quantitative calculations and insights. There is a long history
associated with this small pair theory.  Leggett\cite{Leggett} first
proposed a variational ground state wave function which exhibited a
smooth interpolation between the BCS and Bose-Einstein regimes. An
essential component of his approach was the introduction of a self
consistent equation for the chemical potential to be solved in
conjunction with the variational conditions \cite{eagles}. This $\mu$
differs significantly from the Fermi energy in all but the very weak
coupling limit.  Nozieres and Schmitt-Rink (NSR) \cite{NSR} extended
Leggett's ideas to non-zero temperature and thereby deduced a
transition temperature $T_c$ which varied continuously with increasing
coupling constant $g$ from the BCS exponential dependence on $g$ to
the Bose-Einstein asymptote. Randeria and co- workers\cite{Randeria}
were among the first to apply the NSR approach, derived via a saddle
point scheme, to the cuprates, which were associated with intermediate
values of $g$.  Subsequent extensive numerical\cite{numerics} and
analytical\cite{crossoverothers} studies have addressed the
intermediate $g$ regime beyond the NSR approximations and find a
variety of aspects of pseudogap behavior.

An important goal of the present paper is to address the cross-over
problem within a specific diagrammatic scheme, which is amenable to
quasi-analytic calculations and physical interpretation. In the
process we are able to compute the electronic self energy, spectral
function and a quantitative phase diagram for $T_c$ and $T^*$ as a
function of $g$.  We refer to this decoupling scheme as the ``pairing
approximation'' theory. It is based on early work by Kadanoff and
Martin\cite{Kadanoff} and has been argued by these authors to be the
most appropriate treatment for introducing pairing correlations of the
kind encountered in a conventional superconducting state. Moreover,
this approximation has been extensively applied by
Patton\cite{Patton,GGo} to the problem of $T>T_c$ fluctuations in low
dimensional, dirty superconductors. As will be discussed in detail in
a future paper\cite{Ioan} which addresses the superconducting state,
the present diagram scheme leads naturally to Leggett's\cite{Leggett}
description of the interpolated ground state.  This is in contrast to
alternative diagrammatic approaches
\cite{Haussmann,SereneNSR,Serene,Engelbrecht,GG} which, below $T_c$,
do not recover the BCS limit in the weak coupling regime.

As a result of these calculations, a simple physical picture of the
pseudogap phase emerges. This is a regime of intermediate coupling $g$
which is characterized by a normal state that is somewhere between the
``free'' fermions of the BCS limit and the ``pre-formed'' pairs of the
Bose-Einstein case. A natural description of the interpolated state is
to presume that it corresponds to ``resonantly'' scattered or
meta-stable pairs. Indeed, our calculations of the electronic self
energy and spectral function\cite{Janko} show that the presence of
resonant scattering (as deduced from the character of the pair
propagator or T-matrix), correlates rather well with the break-down of
the Fermi liquid phase. Once the pairs are meta- stable they block
available states around the Fermi surface which would otherwise be
occupied by single electrons and thereby introduce a pseudogap into
the electronic spectrum.

While this interpretation of the density of states depression is in
the same spirit as earlier work\cite{Patton} on superconducting
fluctuation induced spectral gaps, the present description differs in
an important way. Long lived pair states in ordinary superconductors
occur in the vicinity of $T_c$ as a consequence of critical slowing
down.  By contrast, here they are found to occur at much higher
temperatures simply because the coupling $g$ is sufficiently strong to
bind the pairs into a meta-stable state. This picture must necessarily
be distinguished from the phase fluctuation picture of Emery and
Kivelson\cite{Emery} as well as from the ``pre-formed'' pair models of
other groups\cite{Micnas,Ranninger,Alexandrov,TDLee,Tchern}.  Our
pairs correspond to microscopic as distinct from mesoscopically
established regions of superconductivity. Moreover, the resonant pairs
of the present model have a finite lifetime and spatial extent and do
not obey Bose statistics.

\section{Theoretical Framework}
\label{Section_2}

We consider a generic Hamiltonian consisting of fermions in a three
dimensional, jellium gas in the presence of an s-wave attractive
interaction, $V_{\bf k,k'} = g \varphi_{\bf k}\varphi_{\bf k'}$, where
$\varphi_{\bf k} = (1+k^2/k^2_0)^{-1/2}$ and $g<0$ is the coupling
strength expressed in units of $g_c=-4\pi/mk_0$.  Effects associated
with quasi two dimensional lattices and d-wave pairing have also been
addressed within this formalism and will be discussed in a companion
paper\cite{Qinjin}.  The results which we present here are generally
robust, although the relevant energy scales decrease considerably as
the system becomes more two dimensional. Because we consider only this
simplest case, we defer the discussion of direct comparison with ARPES
data to future work.

Following earlier work\cite{Kadanoff,Patton} by Kadanoff and Martin
and by Patton, the system can be characterized by the integral
equations of the ``pairing approximation'' which describe a coupling
between the electronic self energy $\Sigma$ and T matrix $t$. We add
to these equations the usual expression for particle number
conservation
\begin{mathletters}
\label{eq:1}
\begin{eqnarray}
\Sigma^{ }_{{\bf k},i\omega_l} & \;=\; & T\sum_{{\bf q},\Omega_m}
t^{ }_{{\bf q},i\Omega_m}G^{(0)}_{{\bf q-k},i\Omega_m-i\omega_l}
\varphi^2_{{\bf k-q}/2} \label{Sigma} \;,\\
t^{-1}_{{\bf q},i\Omega_m} & = & g^{-1} + T\sum_{{\bf k},\omega_l}
G^{ }_{{\bf k},i\omega_l}
G^{(0)}_{{\bf q-k},i\Omega_m-i\omega_l}
\varphi^2_{{\bf k-q}/2} \label{T-matrix} \;,\\
n & = & \frac{1}{\beta}\sum_{ {\bf k},\omega_l,\sigma} G^{ }_{{\bf 
k},i\omega_l}\;.
\label{n}
\end{eqnarray}
\end{mathletters}
where the Green's function is given by $G^{-1}_{{\bf k},i\omega_l} =
G^{(0) -1}_{{\bf k},i\omega_l} - \Sigma^{ }_{{\bf k},i\omega_l}$, the
bare propagator is $G^{(0) -1}_{{\bf k},i\omega_l} =
i\omega_l-\epsilon_{\bf k}$, and $\Omega_m/\omega_l$ are the even/odd
Matsubara frequencies, with electronic dispersion given by
$\epsilon_{\bf k}=k^2/2m-\mu$.

\begin{figure}
\narrowtext \epsfxsize=3.3in
\epsfbox{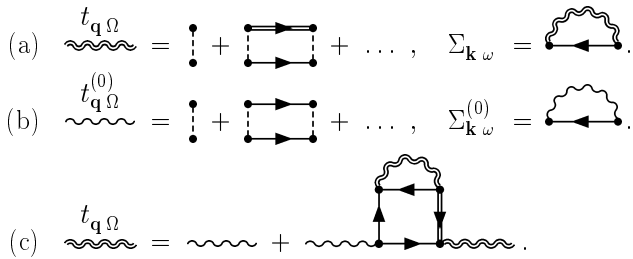}\vspace{- 0.1in}
\caption{The class of diagrams considered in this paper for the
T-matrix and self-energy: (a) The pairing approximation; (b) the
lowest order theory; (c) Dyson's equation for the T-matrix.}
\label{fig:0}
\end{figure}

These first two equations may be readily written diagrammatically as
shown in Figure ~\ref{fig:0}.  For completeness we also indicate the
diagrams of the ``lowest order theory", Figure ~\ref{fig:0}(b), which
will be used initially to build intuition.

Although, these integral equations must in general be solved
numerically, it will be possible to arrive at considerable analytic
insight.  The goal of this paper is to solve these three coupled
equations as a function of arbitrary coupling $g$ for two different
temperature regimes. Our calculations will accordingly proceed in two
stages.  First we discuss the ``lowest order'' conserving theory
(Section ~\ref{Lowest}) where all Green's functions in
Eqs.(\ref{eq:1}) are replaced by bare propagators\cite{Janko}. This
theory is adequate for describing the break down of the Fermi liquid
at $T^*$ and for higher temperatures. In the next stage we (Section
~\ref{MC}) introduce feedback or ``mode coupling" effects in which the
pair propagator is allowed to depend on the single particle self
energy and vice versa. This feedback is necessarily important within
the pseudogap phase and all the way down to $T_c$.  For a range of
temperatures at and above $T_c$ but less than $T^*$, where the
pseudogap is still well established, feedback or mode coupling effects
can be readily introduced via a simple parameterization of the self
energy. We will see below that this simplicity occurs in large part
because of the presence of one ``bare'' Green's function in the first
two equations above.  This parameterization will be justified
numerically in considerable detail.  By contrast, a direct numerical
attack on the problem using the three full equations is difficult to
implement as well as assess because of restrictions to a finite number
of Matsubara frequencies\cite{Schmalian}.  Moreover, it does not
provide the same degree of physical insight, nor can it be done in as
controlled a fashion.

The present scheme should be contrasted with two alternative
approaches in the literature.  The approach of Nozieres and Schmitt-
Rink\cite{NSR} is embedded\cite{SereneNSR} in the limit of Eqs.
(\ref{eq:1}) which corresponds to the lowest order theory and in which
also the right hand side of the number equation is approximated by the
first two terms in a Dyson expansion
\begin{equation}
G  =  G^{(0)} + G^{(0)}\Sigma^{(0)} G^{(0)}
\label{DysonNSR}
\end{equation}
As a consequence of this last approximation there are general problems
with the violation of conservation laws which can be immediately
corrected by summing the full Dyson equation for $G$.  In this way our
conserving ``lowest" order theory is obtained as a first stage
improvement over the NSR scheme. Along related lines, the
Baym-Kadanoff criteria have been applied\cite{wecheckedn,Patton} to
verify conservation laws in the context of the full, mode coupled
theory in Eqs. (\ref{eq:1}).  Finally, Ward identities can be imposed
in conjunction with these full equations to arrive at a satisfactory
description of two particle properties as has been extensively
investigated in Ref.~\cite{Patton}.

Much attention has been paid to an alternative scheme in which all
bare Greens functions in Eqs. (\ref{eq:1}) are replaced by their
dressed counterparts.  This approach, often referred to as the
fluctuation exchange scheme or FLEX, has been studied by direct
numerical methods in finite size systems\cite{GG} as well as for the
purposes of computing $T_c$\cite{Haussmann}.  While it is a ``$\Phi$
derivable'' theory in the sense of Baym\cite{Baym} it was noted some
time ago\cite{Kadanoff} to be inconsistent with BCS theory, even in
the weak coupling limit.  It is, consequently, useful to explore the
present Kadanoff-Baym-Patton approach both as a point of comparison
for results obtained within the FLEX scheme, and because it appears to
be more analytically tractable and more readily interpreted within the
context of conventional theories of superconductivity.

\section{The Lowest Conserving Order}
\label{Lowest}

The single particle self-energy, $\Sigma_{{\bf k},\omega}$, at the
lowest conserving level of approximation corresponds to
\begin{equation}
\Sigma^{(0)}_{{\bf k},i\omega_l}  = T\sum_{{\bf q},\Omega_m}
t^{(0)}_{{\bf q},i\Omega_m}G^{(0)}_{{\bf q-k},i\Omega_m-i\omega_l}
\varphi^2_{{\bf k-q}/2}\, ,
\label{self-energy}
\end{equation}
and the T-matrix is given by
\begin{equation}
t^{(0) -1}_{{\bf q},i\Omega_m} = g^{-1} + T\sum _{{\bf p},\omega_l}
G^{(0)}_{{\bf p},i\omega_l}
G^{(0)}_{{\bf q - p},i\Omega_m - i\omega_l}
\varphi^2_{\bf p-q/2}\, .
\label{T-matrix0}
\end{equation}
These equations are indicated diagrammatically in Figure 
~\ref{fig:0}(b).

Although this formulation of the problem will not give quantitatively
correct values for properties such as the transition temperature, it
does yield qualitatively correct results for the self-energies and
spectral functions. Indeed pseudogap effects in the spectral
properties of the system are produced already at this
order\cite{Janko}.  This approach becomes increasingly more reliable
in the weak pseudogap regime, where feedback effects are expected to
be negligible.

In the next two subsections we discuss and plot the behavior of the
T-matrix $t^{(0)}_{{\bf q}, \Omega}$ and, with these insights, then
address the electronic self energy $\Sigma^{(0)}_{{\bf k},\omega}$
which is determined from an integral over the T-matrix.  Fundamental
to our considerations is the observation that for sufficiently strong
coupling, when the fermions are bound into composite bosons, a Fermi
liquid description is clearly inappropriate.  The goal of the present
discussion is to determine for what values of $g$ does this breakdown
of the (normal state) Fermi liquid happen and does this occur well
before the system is in the strict ``preformed pair" limit?  In a
related fashion, we address the temperature dependence of this
breakdown, for a given moderately large $g$ value. In this way, the
lowest order scheme will allow us to establish a criterion for
determining $T^*$.

\subsection{T Matrix: From Weak to Strong Coupling}

 The lowest order T-matrix  can be written as
\begin{equation}
t^{(0) -1}_{{\bf q},i\Omega_m} = g^{-1}_{ } + \sum_{\bf k}
\frac{1-f(\epsilon_{\bf k})-f(\epsilon_{\bf k-q})}
{\epsilon_{\bf k}+\epsilon_{\bf k-q}-i\Omega_m}\,\varphi^2_{\bf k-
q/2}\, ,
\label{InT-normal}
\end{equation}

The imaginary component of the electronic self energy may be directly
computed via
\begin{eqnarray}
{\it Im}\,\Sigma _{{\bf k},\omega} =  \sum_{\bf q}\
\bigl[b(\omega + \epsilon _{\bf q-k}) + f(\epsilon_{\bf q-
k})\bigr]\nonumber\\
\ \times {\it Im}\,t_{{\bf q},\omega + \epsilon _{\bf q-k}}
\varphi^2_{\bf k-q/2}\, .
\label{ImSigma}
\end{eqnarray}
where $f(x),b(x) = (e^{\beta x} \pm 1)^{-1}$.  (and the real part
obtained via Kramers Kronig transforms). This expression, which is not
restricted to the lowest order theory, will be used throughout the
paper.

\begin{figure}
\narrowtext
\epsfxsize=3.3in 
\epsfbox{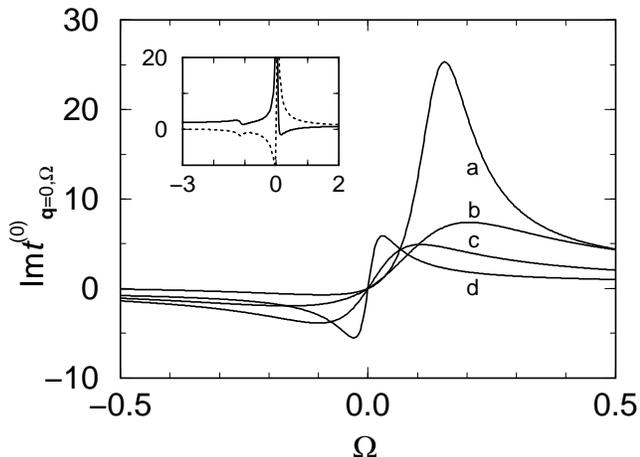}
\vspace{-0.1in}
\caption{{\bf Main figure}: Frequency dependence of ${\it
Im}\,t^{(0)}_{{\bf q = 0},\Omega }$ for $T/T_c = 1.5$ and coupling (a)
$g/g_c$ = 1.5; (b)$g/g_c$ = 1.2; (c) $g/g_c$ = 1.0; (d) $g/g_c$ =
0.6. {\bf Inset}: Real (solid) and imaginary (dashed) components of T
matrix over extended frequency range, with $g/g_c = 1.2$ and $T/ T_c =
1.1$.In this and all subsequent figures energies are measured in units
of $E_F$.}
\label{fig:1}
\end{figure}

It can be seen from Eq. (\ref{ImSigma}) that the self energy is
determined by ${\it Im}\,t_{{\bf q},\Omega }$.  It is instructive to
plot this quantity in the lowest order theory, as shown in Figure
\ref{fig:1}, for a range of coupling constants and at fixed relative
temperature $T/ T_c = 1.5$, with wave-vector ${\bf q} = 0 $ .  Here
$T_c$ is obtained from the usual Thouless criterion for the divergence
in the T matrix $t^{(0)}_{{\bf q = 0 },\omega = 0 }$.  The inset shows
the full analytic T matrix function on the real axis over a more
extended frequency range and for fixed coupling $g/g_c = 1.2$ and $T/
T_c = 1.1$. It can be seen from the main figure that the major effect
of increased coupling is to introduce a finite frequency peak
structure into ${\it Im}\,t_{{\bf q},\Omega }$ whose strength grows
with $g/g_c$.  Using Eq. (\ref{ImSigma}), it follows that this peak
structure must reflect itself in the self energy.  However, on the
basis of these plots there is no obvious dividing line associated with
a $\it{qualitative}$ change in $\it {Im}\Sigma$.  Such a qualitative
change, which signals the breakdown of the Fermi liquid state, is
expected, somewhere between the regime of weak and strong coupling.

\begin{figure}
\narrowtext \epsfxsize=3.3in
\epsfbox{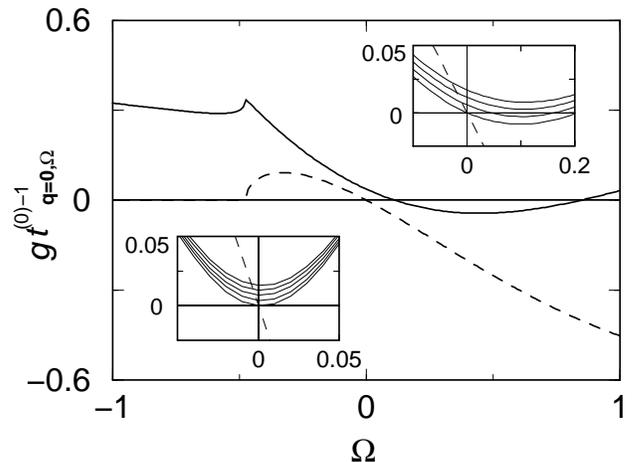}\vspace{-0.1in}
\caption{{\bf Main figure}: Resonance effects illustrated by real
(solid) and imaginary parts (dashed line) of $g t^{(0) -1}_{{\bf q =
0},\omega}$, for $g/g_c = 1.2$, and $T/T_c = 1.1$.  {\bf Upper inset}:
${\rm Re}\, g t^{(0) -1}_{{\bf q = 0},\omega} $ (solid curves) for $T/T_c$
from $1.0$ (bottom curve) to $1.1$ (top curve), and $g/g_c = 1.2$.
Dashed curve is ${\rm Im}\, g t^{(0) -1}_{{\bf q = 0},\omega}$, which is
unchanged for these $T/T_c$.  {\bf Lower inset}: Same as in upper
inset with $g/g_c =0.6$.  }
\label{fig:2}
\end{figure}

To identify more precisely what is the nature of those effects in the
T-matrix which are responsible for the Fermi liquid breakdown, it is
instructive to plot the real and imaginary parts of the $\it{inverse}$
function $t^{(0) -1}_{{\bf q},\omega}$ which is shown in Figure
\ref{fig:2} for moderately strong coupling $g/g_c = 1.2$, and
temperature $T/T_c = 1.1$. Precisely at $T_c$ both the real and
imaginary parts of the inverse T matrix at $ {\bf q} = 0, \omega = 0$
are zero. However at $T/T_c = 1.1 $ and for ${\bf q} = 0$ it can be
seen that the real part of the inverse T matrix is zero at finite
frequency, corresponding to a pair $\it{resonance}$.  Moreover, this
resonance disperses with wave vector ${\bf q} $. It should be noted
that there is a second zero crossing of the real part of the inverse
T-matrix which is not associated with a resonance, since for this
higher frequency the imaginary part of the inverse T-matrix is large.

As shown in the upper right inset, for temperatures $T/T_c$ ranging
from $1.0$ to $1.1$, (and for the case $g/g_c = 1.0$), as the
temperature is raised, the resonance gradually disappears.  This
behavior is qualitatively different for weaker coupling ( $g/g_c =
0.6$) as shown in the lower left inset. In this latter case the only
time the real part of the inverse T-matrix is zero is in the critical
regime, which essentially coincides with $T_c$.  The characteristic
dividing line between resonant and non-resonant behavior appears to be
around $g/g_c \approx 1.0$, for this value of the interaction range.
For slightly different parameterizations of the interaction this
number may change slightly.  We will show in the next subsection that
these resonance effects lead to non Fermi liquid like behavior in the
electronic self energy.

\subsection{Self Energy: From Weak to Strong Coupling}

To confirm our numerical results, we will compute the self energy
$\Sigma^{(0)} _{{\bf k},\omega}$ in two different ways.  ${\it
Im}\,\Sigma_{{\bf k},\omega}$ is derived by direct evaluation of the
real frequency integral and the Kramers-Kr\"onig relation used to
obtain ${\it Re}\,\Sigma_{{\bf k},\omega}$. We refer to this as the
direct real--axis calculation.  Alternatively, we consider the
self-energy at the Matsubara points using
\begin{equation}
\Sigma_{{\bf k},i\omega_l} = \sum_{\bf q}\int^{\infty}_{-\infty}
\frac{{\rm d}\Omega}{\pi}\ b(\Omega)\ \frac{{\rm Im} t^{ }_{{\bf q},\Omega}\,
\varphi^2_{\bf k-q/2}} {\Omega-i\omega_l-\epsilon_{\bf k-q}}\, .
\label{Sigma-nof}
\end{equation}
It should be noted that this expression is equivalent to ignoring the
Fermi term $ f(\epsilon_{\bf q-k})$ in Eq. (\ref{ImSigma}). This term
is {\em not} important in the well established pseudogap phase, but it
is necessary to recover Fermi liquid behavior.  A Pad\'e approximant
scheme \cite{Vidberg} is then used to analytically continue the
self-energy to the real frequency axis.  These two approaches -- the
direct real--axis calculation, and the Matsubara scheme -- gave rather
similar results away from the Fermi liquid regime, although only the
former led to a proper Fermi liquid signature in the self energy.  The
second of these two schemes was used in earlier work by our
group\cite{Janko}.

\begin{figure}
\narrowtext \epsfxsize=3.3in
\epsfbox{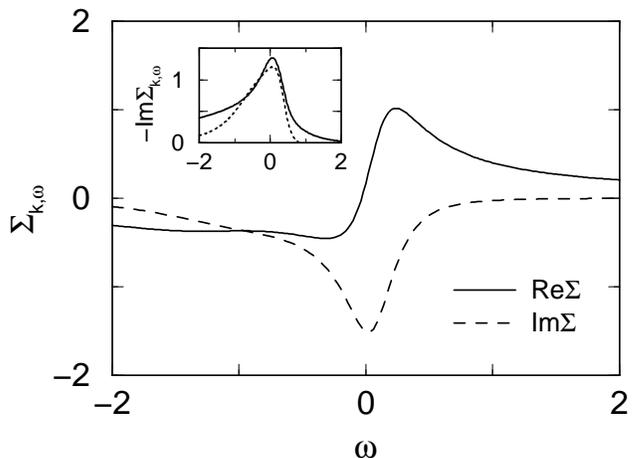}\vspace{-0.1in}
\caption{{\bf Main figure}: ${\rm Re} \Sigma_{{\bf k_F},\omega}$
(solid line), and ${\rm Im} \Sigma_{{\bf k_F},\omega}$ (dashed line),
within the lowest order theory. {\bf Inset} shows a comparison between
results for ${\rm Im} \Sigma_{{\bf k_F},\omega}$ using direct real
axis (solid ) and Matsubara (dashed line) schemes. }
\label{fig:3}
\end{figure}

In Figure \ref{fig:3} we plot the complex self energy function
obtained from the Pad\'e approximant approach for $g/g_c =
1.2$. For these intermediate coupling strengths, a resonance
condition in the T-matrix leads to a peak in $-{\it Im}\,\Sigma_{{\bf
k},\omega}$ at $\omega + \epsilon_{\bf k} \approx 0$. This is in
contrast to the Fermi liquid regime where $-{\it Im}\,\Sigma_{{\bf
k},\omega}$ exhibits a minimum rather than a maximum. This peak
reflects that in the T-matrix, however, further amplified by the zero
frequency divergent contribution of the Bose factor. These
observations are confirmed by the real--axis calculations as well; the
two results for $-{\it Im}\,\Sigma_{{\bf k},\omega}$ are compared in
the inset and agree reasonably well.

\begin{figure}
\narrowtext \epsfxsize=3.3in
\epsfbox{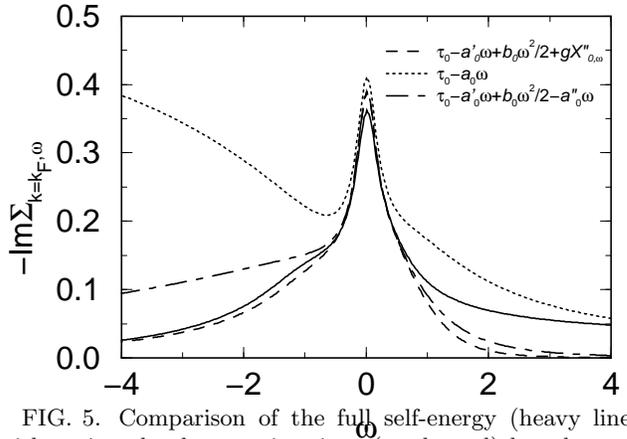}\vspace{-0.3in}
\caption{Comparison of the full self-energy (heavy line), with various
level approximations (see legend) based on expanding $g t^{(0)
-1}_{{\bf q},\omega}$ at low ${\bf q } , \Omega$. }
\label{fig:4}
\end{figure}

The evident association between a peak in $-{\it Im}\Sigma^{(0)}
_{{\bf k},\omega}$ and a resonance condition in the corresponding T
matrix ($t^{(0) -1}_{{\bf q},\Omega}$) suggests that it is primarily
the resonance structure ( at low ${\bf q}$ and low $\Omega$ ) in the
latter which determines the behavior of the self energy. To check this
conjecture, and to provide some simplification for future
calculations, in Figure ~\ref{fig:4} we compare $-{\rm Im}
\Sigma^{(0)} _{{\bf k},\omega}$ deduced using various Taylor series
expansions in ${\bf q}$ and $\Omega$ for the inverse T-matrix . The
result obtained from the full expression (\ref{InT-normal}) is
indicated for comparison by the solid curve \cite{q=0}.  Reasonable
agreement is obtained provided a sufficient number of terms (in
$\Omega$) are included in this Taylor series or time dependent
Ginzburg- Landau (TDGL) expansion of the T- matrix
\begin{equation}
gt^{LG -1}_{{\bf q},\Omega} = \tau_0 - (a'_0+ia''_0) \Omega +
b_0\Omega^2 + \xi^2_{\rm LG} q^2\, .
\label{LG-exp}
\end{equation}
Here the parameters $a'_0$, $a''_0$, $b_0$, and $\xi^2_{\rm LG}$ are
calculated by expanding the full expression for $t^{(0)}_{{\bf
q},\Omega}$. Because of this reasonably good agreement, and because of
the numerical simplicity it introduces, in the remainder of this
section the direct real axis self energy calculations are based on
this Landau- Ginzburg form. For our Pad\'e calculations, we included
the full $\Omega, {\bf q }$ dependences.

It should be stressed that the calculations in the previous subsection
demonstrate that the full T-matrix $t^{(0) }_{{\bf q},i\Omega} $ has a
complicated analytic structure. Thus it should never be assumed that
this function is equivalent to $t^{LG}_{{\bf
q},\Omega}$. Nevertheless, the above calculations of $-{\it
Im}\Sigma^{(0)} _{{\bf k},\omega}$ indicate that, for the purposes of
computing the self energy, it is adequate to introduce a simplified
form for the T-matrix.

The quadratic in $\Omega$ terms of Eq. (\ref{LG-exp}) were found to be
reasonably important for obtaining agreement in the tails of the self
energy function. These quadratic terms capture the physics of the
second zero crossing of the inverse T-matrix, shown in the upper right
hand inset of Figure \ref{fig:2}.  We will see below that very near
$T_c$ most of the physics is dominated by the low frequency regime so
that $t^{LG}_{{\bf q},\Omega}$ is well approximated by a simple pole
structure

\begin{equation}
t^{resonant}_{{\bf q},\Omega} = \frac{- g(a'_0)^{-1}}{\Omega-
\Omega_{\bf q}+i\Gamma_{\bf q}} ,
\label{T-resonant1}
\end{equation}

where $\Omega_{\bf q} = (\tau_0 + \xi^2_{\rm LG} {\bf q}^2)/a'_0$, and
$\Gamma_{\bf q} = \Omega_{\bf q} (a''_0/a'_0)$. Eq. (\ref{T-resonant1}), thus,
reflects that component of the T-matrix which dominates the self
energy and related quantities near $T_c$.

\begin{figure}
\narrowtext \epsfxsize=3.3in
\epsfbox{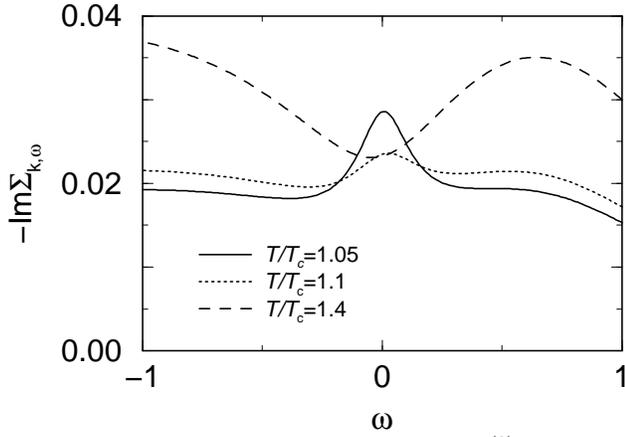}\vspace{- 0.1in}
\caption{Temperature dependence of $-{\it Im}\Sigma^{(0)} _{{\bf
k_F},\omega}$ as a function of $\omega$, for $T/T_c =$ 1.05, 1.1 and
1.4.  As the temperature is lowered, the quadratic minimum at zero
frequency -- the salient feature of a Landau Fermi liquid -- is
gradually turned into a maximum in the pseudogap regime.}
\label{fig:5}
\end{figure}

We end this subsection by studying the evolution with temperature of
$-{\it Im}\Sigma^{(0)} _{{\bf k},\omega}$ as a function of $\omega$
for $g/g_c = 1.2$.  The results are plotted in Figure \ref{fig:5}
where the three curves are at different temperatures (given
respectively by 1.05, 1.1 and 1.4 $T/T_c$) and correspond to
temperatures just below, at and significantly above the onset for pair
resonance, which occurs at approximately $1.2 T_c$.  A Fermi liquid
like minimum is present only for the high temperature regime. This
behavior mirrors the evolution with coupling constant seen at fixed T.
It should be stressed that, upon increasing $T$, when a single peaked
spectral function is first recovered, this peak is still quite
broad. This phase should, thus, not be associated with a canonical
Fermi liquid.
 
\subsection{Parameterization of $\Sigma$ and Spectral Function Gap}
\label{Param}

In this subsection we identify a pseudogap parameter and associate
this parameter with the calculated gap in the spectral function. In
the process we arrive at a convenient parameterization of the self
energy which will be useful later on. This parameterization works
better as temperature decreases towards the well established pseudogap
regime. Thus this parameterization forms the foundation for later work
within the full mode coupling scheme.

\begin{figure}
\narrowtext \epsfxsize=3.3in
\epsfbox{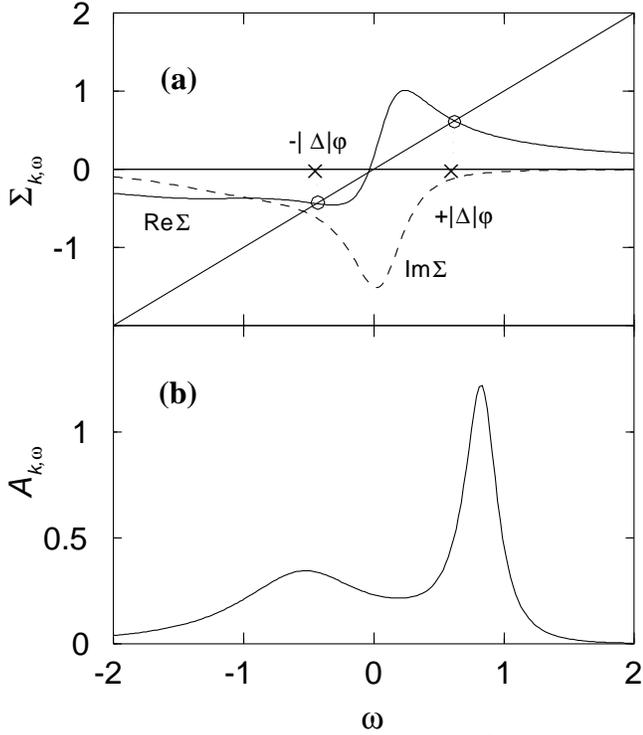}
\caption{(a) Frequency dependence of $\Sigma^{(0)} _{{\bf
k_F},\omega}$. The single particle spectral function will have a peak
(at $\omega$ indicated by the circles) whenever $ {\rm Re}
\Sigma_{{\bf k_F},\omega}$ crosses the unit diagonal (shown by the
line) and $ {\rm Im} \Sigma_{{\bf k_F},\omega}$ is small.  (b)
Associated spectral function $A_{{\bf k}, \omega}$ obtained from the
self-energy plotted in (a). }
\label{fig:6}
\end{figure}

In Figure \ref{fig:6}(a) we re-plot the self energy of Figure
\ref{fig:3} along with the straight line curve for $\omega$. The
points at which the line intersects $-{\it Re}\,\Sigma_{{\bf
k},\omega}$ are important because they lead to structure in the
Green's function $G_{{\bf k},\omega}$.  In this way we determine where
the real part of the inverse Green's function, ${\it Re}\,G^{-1}_{{\bf
k},\omega}=\omega-\epsilon_{\bf k}-{\it Re}\,\Sigma_{{\bf k},\omega} =
0 $, corresponding to a set of three zeros. Since the imaginary part
is small for the two outer roots, the spectral function $A_{{\bf
k},\omega}$, which is given by $A_{\bf k} (\omega ) = -\pi^{-1}{\rm
Im}\,[G_{\bf k} (i\omega _l \rightarrow \omega +i0)]$, acquires two
peaks separated by $2|\Delta_{pg}| \varphi_{\bf k}$ with
\begin{equation}
|\Delta_{pg}|^2 \approx - \sum_{\bf q}\int^{+\infty}_{-
\infty}\frac{{\rm d}\Omega}{\pi}\, b(\Omega)\, {\it 
Im}\,t^{(0)}_{{\bf
q},\Omega}\, .
\label{Delta}
\end{equation}
Note that within our sign conventions $ |\Delta_{pg}|^2 $ is a
positive definite quantity. The spectral function is plotted in Figure
\ref{fig:6}(b).  The gapped structure thus corresponds to a pseudogap
in the electronic properties ( such as the density of states). This
should be contrasted with the single peak in $A_{{\bf k},\omega}$
associated with the Fermi liquid phase (at weaker coupling or higher
temperatures). The asymmetric broadening of the two peaks is a generic
feature of this pseudogap model and is due to the interaction of
correlated pairs with the occupied states of the Fermi sea.  The
dispersion of the two peaks has been plotted elsewhere\cite{Janko}.
As the momentum vector ${\bf k}$ passes through the Fermi surface the
spectral weight shifts from the negative to the positive frequency
peak.  Close to the Fermi momentum the peaks disperse roughly as
$E_{\bf k}=\pm\sqrt{\epsilon_{\bf k}^2+|\Delta_{pg}|^2\varphi^2_{\bf
k}}$.

\begin{figure}
\narrowtext \epsfxsize=3.3in
\epsfbox{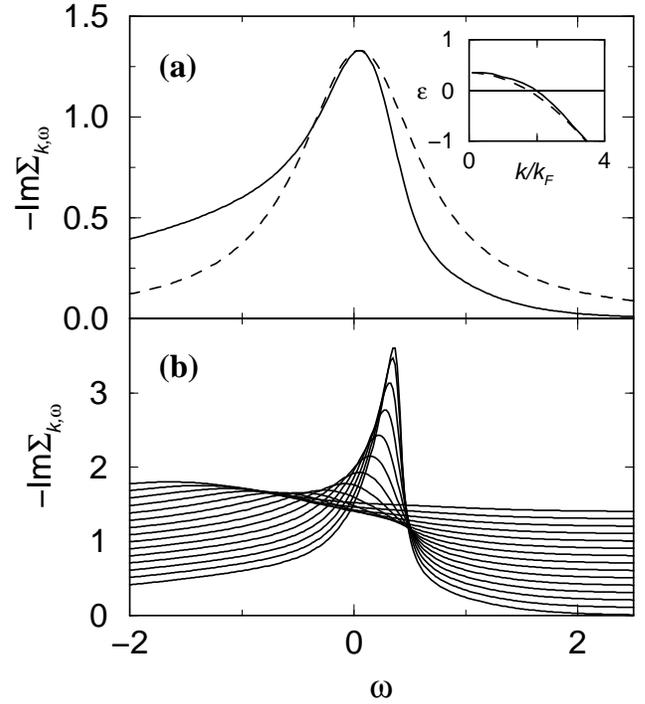}\vspace{- 0.1in}
\caption{(a) {\bf Main figure}: Numerically calculated ${\it -
Im}\,\Sigma_{{\bf k},\omega}$ (heavy line) and Lorentzian fit (dashed
line) parameterized by Eq. (\ref{Sigma-model}). {\bf Inset}: The
dispersion of peak in full (solid line) and model (dashed line) self
energy; (b) ${\it - Im}\,\Sigma_{{\bf k},\omega}$ for $k/k_F = 0.1$
(bottom curve) to $k/k_F = 1.5$ (top curve) . The curves are
vertically offset for clarity.}
\label{fig:7}
\end{figure}

Given the characteristic peak in ${\it Im}\,\Sigma_{{\bf k},\omega}$,
it is useful to consider a two parameter model self energy of the 
form
\cite{newfootnote}
\begin{equation}
\Sigma^{model}_{{\bf k},\omega} \approx 
\frac{|\Delta_{pg}|^2\varphi^2_{\bf k}}{\omega+\epsilon_{\bf k}
+i\gamma}\, ,
\label{Sigma-model}
\end{equation}
A fit of ${\it Im}\,\Sigma_{{\bf k},\omega}$ to
Eq. (\ref{Sigma-model}) is presented in Figure \ref{fig:7}(a).  Here
the adjusted parameter is $\gamma$ since $\Delta_{pg}$ can be
evaluated from Eq. (\ref{Delta}).  The major difference between the
two curves is evident primarily in the tails of the functions. As the
temperature gets closer to $T_c$, we will see that the model becomes
progressively better.

In order to further assess the utility of the model, it is instructive
to examine the dispersion of the peak of ${\it Im}\,\Sigma_{{\bf
k},\omega}$.  Figure ~\ref{fig:7}(b) plots ${\it Im}\,\Sigma_{{\bf
k},\omega}$ for a series of wave vectors. This peak dispersion can be
compared with that of the model, as is shown in the inset of Figure
\ref{fig:7}(a).  Agreement is reasonable, and on this basis we will
make use of Eq. (\ref{Sigma-model}) in later studies.

\section{Mode coupling theory}
\label{MC}

In this section we address the three coupled integral equations shown
in Eqs. (\ref{eq:1}) in the full mode coupled regime.  These equations
must be solved numerically.  We have just seen that the numerics are
tractable for temperatures at and above $T^*$, where the ``lowest
order" theory is adequate.  The transition from the Fermi liquid
breakdown to the well established pseudogap phase is difficult to
treat and its detailed characterization will be deferred to a future
paper. From the well-established pseudogap state, down to $T_c$, we
will show that the numerics again become tractable. We first address
the T-matrix and self energy in this regime, concentrating on
moderately strong coupling ($g/g_c = 1.2$), where there is an
appreciable pseudogap.  We next use these results to determine the
behavior of $T_c$ as a function of arbitrary $g$.  This latter
quantity can be derived from the Thouless criterion or T-matrix
divergence.  Mode coupling effects enter in an important way, since
the T-matrix differs from that deduced in the lowest order theory. The
superconductivity that results within the mode coupling scheme should
be viewed as deriving from pairing of pseudogapped electrons. This
superconductivity is, thus, not associated with a Fermi liquid state.

\subsection{Behavior of $\Sigma$ and the T-matrix near $T_c$}

The coupled equations for $\Sigma_{{\bf k},\omega}$ and $t^{- 1}_{{\bf
q},\Omega}$, as a function of temperature, were solved numerically
using the model self-energy of Eq. (\ref{Sigma-model}).  In a given
iteration $\Delta_{\rm pg}$ and $\gamma$ were used to compute $t_{{\bf
q},\Omega}$ via Eq. (\ref{T-matrix}). The result was then substituted
into Eq. (\ref{Sigma}).  The output of the latter was then used to
extract $\Delta_{\rm pg}$ and $\gamma$ via (least squares) fits to
Eq. (\ref{Sigma-model}) and the procedure repeated until convergence.
Once self--consistency in these parameters is reached, the resulting
self-energy and T-matrix can be calculated as an output from the {\em
full} equations (\ref{Sigma}-\ref{T-matrix}). To distinguish these
from the modeled functional forms, in the following
we refer to these results as the {\em full functions} for
${\it Im}\,t_{{\bf q},\Omega }$ and ${\it Im}\Sigma_{{\bf 
k},\omega}$. Unless indicated otherwise, these are the quantities 
plotted
throughout the remainder of this paper.

\begin{figure}
\narrowtext \epsfxsize=3.3in
\epsfbox{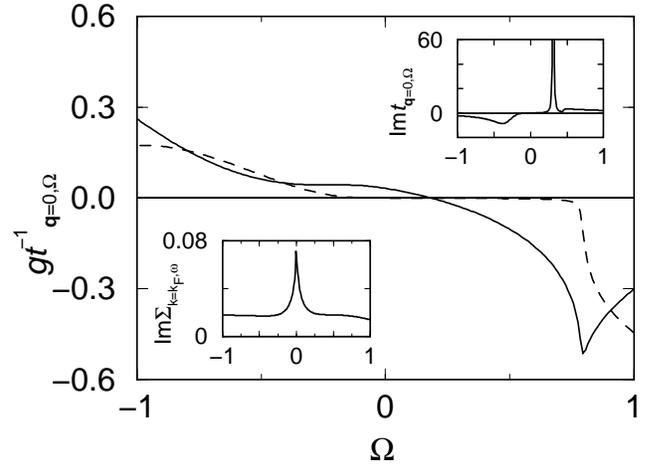}\vspace{- 0.1in}
\caption{{\bf Main figure}: The real (solid) and imaginary part
(dashed line) of inverse T-matrix $g t^{ -1}_{{\bf q},\Omega}$ at full 
mode coupling level, for temperatures slightly above $T_c$.
{\bf Upper right inset} plots ${\it Im}\,t_{{\bf q},\Omega }$ and
{\bf Lower left inset} plots $-{\it Im}\Sigma_{{\bf k},\omega}$ at
the same temperature.}
\label{fig:8}
\end{figure}

The resulting inverse T- matrix $t^{ -1}_{{\bf q},\Omega}$ is plotted
in Figure \ref{fig:8}. The upper right inset plots ${\it Im}\,t_{{\bf
q},\Omega }$ and the lower inset $-{\it Im}\Sigma_{{\bf k},\omega}$
for a temperature slightly above $T_c$.  (Proximity to $T_c$ is
signaled by an incipient divergence in the T-matrix).  Here, $g/g_c =
1.2$. This behavior should be compared with the lowest order results
in Figures \ref{fig:1},\ref{fig:2},\ref{fig:3}.  The salient feature
in the main body of the figure is that a resonance appears, just as in
the lowest order calculations.  What is striking, however, is that the
resonance is sharper as a consequence of feedback effects, due to a
suppressed, essentially gapped, imaginary part.  The (approximate) gap
in the imaginary part is to be directly associated with the pseudogap
in the electronic spectrum.  Thus we have the interesting result that
feedback effects stabilize the resonance found in the lowest order
theory. Stated in more physical terms, as $T_c$ is approached, pairs
of low momenta become very long lived.

The upper right inset shows ${\it Im}\,t_{{\bf q},\Omega }$ which
exhibits a significantly sharper peak structure than was seen in the
lowest order theory. It is to be expected that this peak will reflect
itself in $-{\it Im}\Sigma_{{\bf k},\omega}$, which is plotted in the
lower left inset. It should be noted that the figure plots the full
function not that of Eq. (\ref{Sigma-model}) to which it can be well
fitted. Indeed the fit becomes progressively better as temperature is
lowered, and concomitantly $\gamma$ becomes smaller.

This analysis will be used later to determine the actual temperature
at which the T-matrix diverges. As the temperature is lowered towards
$T_c$ several important simplifications occur: (i) $\gamma$ approaches
zero and (ii) most of the contribution to the self energy is captured
by approximating the T-matrix inside the integral of Eq.
(\ref{ImSigma}) by the TDGL form. Point (i) indicates that the self
energy behavior near $T_c$ is a continuation of the (much weaker) peak
in ${\it Im}\,\Sigma_{{\bf k},\omega}$ at $\omega=-\epsilon_{\bf k}$
which was found earlier as the Fermi liquid breaks down near $T^*$.
With lower T this peak grows progressively sharper until the
instability at $T_c$ is reached. The TDGL expansion noted in (ii) is
given by Eq. (\ref{LG-exp}). This equation reflects the known analytic
form (at low $\Omega$ and ${\bf q}$) of the divergence in the
T-matrix. It is not unexpected that this incipient divergence will
dominate the properties of the electronic self energy. Moreover, this
simplification is readily incorporated into the numerical iteration
scheme discussed above to expedite our calculations.  As an
intermediate step, we fit the calculated T-matrix to
Eq. (\ref{LG-exp}) for use in Eq. (\ref{Sigma}). It should be stressed
that after convergence the full self energy and T- matrix are
produced.  Moreover, at each step the validity of
Eq. (\ref{Sigma-model}) and Eq. (\ref{LG-exp}) are checked by
comparing with their respective counterparts given in
Eq. (\ref{Sigma}) and Eq. (\ref{T-matrix}).

\begin{figure}
\narrowtext \epsfxsize=3.3in
\epsfbox{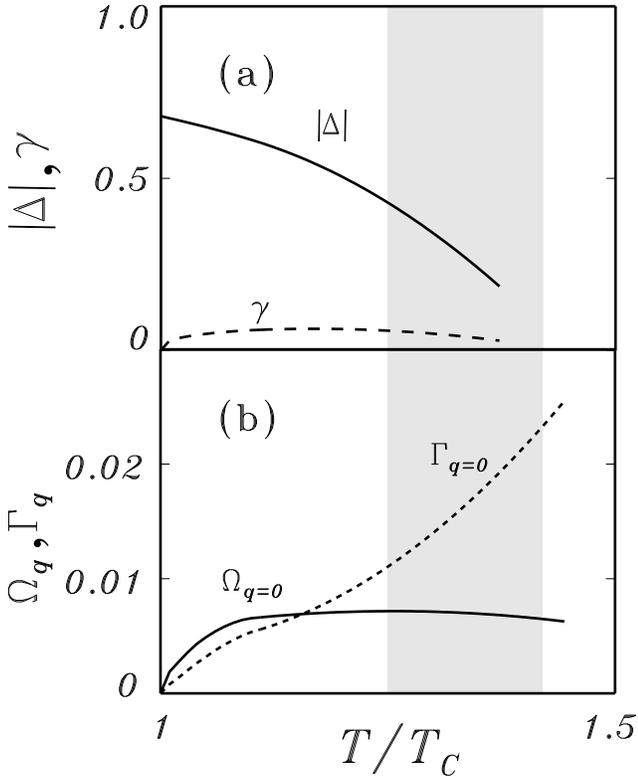}
\caption{The evolution of the single particle parameters 
$\Delta_{\rm
pg}$ and $\gamma$ (a), contrasted with the pair parameters
$\Omega_{\bf q}$ and $\Gamma_{\bf q}$ (b). The coupling is fixed at
$g/g_c = 1.2$. For temperatures above and below the shaded area, 
the lowest order theory and mode coupling scheme are used,
respectively. Curves inside the shaded region were obtained by 
interpolation.}
\label{fig:9}
\end{figure}

We conclude this subsection by comparing the evolution of the single
particle parameters $\Delta_{\rm pg}$ and $\gamma$ with their pair
counterparts which we parameterize as $\Omega_{\bf q}$ and
$\Gamma_{\bf q}$ by fitting the resonance feature in the T-matrix to
Eq.  (\ref{T-resonant1}). The pair parameters, $\Omega_{\bf q}$ and
$\Gamma_{\bf q}$, thus, correspond respectively to the energy and
width of the resonance in the T-matrix. The results of our numerical
iteration scheme are shown in Figure \ref{fig:9} for $g/g_c = 1.2$.
The shaded regions indicate where Eq. (\ref{Sigma-model}) breaks down,
as does the lowest order theory.  Even though the pseudogap persists,
the system begins to cross-over towards the Fermi liquid. Thus the
parameters $\Delta_{\rm pg}$ and $\gamma$ are indicated only for a
limited range of temperatures, where Eq. (\ref{Sigma-model}) is
relevant.  The temperature for Fermi liquid onset $\sim~T^*$, is just
beyond the shaded region corresponding to the regime of validity of
the lowest order theory.  At higher temperatures $\Gamma_{\bf q=0}$
and $\Omega_{\bf q=0}$ are plotted to join, after extrapolation, onto
the results of the lowest order theory. The temperature $T^*$ can be
read off from the endpoint of the two curves in Figure \ref{fig:9}(b).
This is the highest temperature at which $\Gamma$ and $\Omega$ are
defined.

This figure re-enforces the earlier observation that the pairs are
relatively long lived over much of this temperature regime: the decay
rate, $\Gamma_{\bf q=0}$, remains comparable to $\Omega_{\bf q=0}$.
The small value of $\Gamma_{\bf q=0}$ is also reflected in the small
value of $\gamma$.  We may view this (well established pseudogap
phase, to the left of the shaded region in Figure \ref{fig:9}) as an
extended critical regime which arises as a result of the gapping of
the T-matrix via feedback effects.

This variation in $\gamma$, and $\Delta_{\rm pg}$, along with our
earlier analysis from the lowest order theory, (see, for example,
Figure \ref{fig:5}) has direct implications for experiment: the
spectral function $A_{{\bf k},\omega}$ exhibits a single broad peak at
$T > T^*$, which evolves into two peaks at $T \approx T^*$. The
separation between these peaks increases with decreasing temperature,
as $\Delta_{\rm pg}$ increases.  At the same time since $\gamma$
decreases as $T_c$ is approached, the two peaks progressively sharpen
in the vicinity of the superconducting transition. This behavior
represents, then, a smooth transition of pseudogap behavior into the
superconducting state. It should not, however, be assumed that the
pseudogap parameter $\Delta_{\rm pg}$ is equivalent to the
superconducting gap $\Delta_{\rm sc}$ below $T_c$. This latter issue
will be discussed in a future paper\cite{Ioan}.

\subsection{Calculation of $T_c$}

In this subsection we deduce the transition temperature for variable
coupling constant using two different numerical approaches.  We use
the iterative numerical scheme discussed above to determine $T_c$ by
studying the diverging T-matrix as the temperature is lowered. In this
way $T_c$ is obtained as the temperature at which both $\Gamma_{\bf
q=0}$, and $\Omega_{\bf q=0}$ are identically zero.

This scheme yields numerically equivalent answers to those obtained
using an alternative and simpler approach. For the purposes of
calculating $T_c$, the largest contribution to the self energy in
Eq. (\ref{Sigma}), may be seen to come from the low frequency, long
wavelength phase space region where the T-matrix is large. The
integral is well approximated by \cite{newfootnote}
\begin{equation}
\Sigma^{ }_{{\bf k},\omega} \approx -\Delta_{\rm 
pg}^2\,\varphi^2_{\bf
k}\,G^{(0)}_{{\bf k},-\omega} ,
\label{Sigma-Tc} \\
\end{equation}
Indeed, this equation is equivalent to Eq. (\ref{Sigma-model}) in
which $\gamma$ is set to zero, as was found to be the case numerically
at $T_c$.

It follows from this equation that
\begin{equation}
t^{-1}_{{\bf 0},0} = g^{-1} + \sum_{\bf k}\frac{1-2f(E_{\bf
k})}{2E_{\bf k}}\,\varphi^2_{\bf k}.
\label{Thouless-crit}
\end{equation}
where $E_{\bf k} = \sqrt{\epsilon^2_{\bf k}+\Delta_{\rm
pg}^2\varphi^2_{\bf k}}$.  The pseudogap parameter which appears 
in Eq. (\ref{Sigma-Tc}) is given by
\begin{equation}
\Delta_{\rm pg}^2 = - \sum_{\bf q}\int^{\infty}_{-\infty} \frac{{\rm
d}\Omega}{\pi}\, b(\Omega) \,{\it Im}\, t_{{\bf q},\Omega} ,
\label{Delta-def}
\end{equation}
The lowest order analogue of this equation was inferred earlier in
Section 3.  It is evident that $\Delta_{\rm pg}^2$ in
Eq. (\ref{Delta-def}) coincides with the square amplitude of pairing
fluctuations, $g^2 \langle c^{\dagger}c^{\dagger}c^{ }c^{ } \rangle$,
which can be defined more generally away from $T_c$.

In order to evaluate $T_c$, Eq. (\ref{Delta-def}) must be combined
with the Thouless criterion from Eq. (\ref{Thouless-crit}), along with
the number equation. The three fundamental equations for $T_c$ are
Eq. (\ref{Delta-def}), along with
\begin{eqnarray}
1 + g\sum_{\bf k}\frac{1-2f(E_{\bf k})}{2E_{\bf k}}\, \varphi^2_{\bf
k} & = & 0 , \label{Thouless-2}\\ 2\sum_{\bf k}\left[v^2_{\bf k} +
\frac{\epsilon_{\bf k}}{E_{\bf k}}\,f(E_{\bf k})\right] & = & n ,
\label{number}
\end{eqnarray}
where $v^2_{\bf k}=(1-\epsilon_{\bf k}/E_{\bf k})/2$.

It should be noted that Eqs. (\ref{Sigma-Tc}), (\ref{Thouless-2}) and
(\ref{number}) show that the single particle Green's function, the
Thouless criterion and the number equation assume a particularly
simple form which coincides which their (below $T_c$) counterparts
obtained in the standard superconducting theory but with the pseudogap
playing the role of the superconducting gap.  This simplicity (as well
as the related BCS-like behavior, at small $g$, below\cite{Kadanoff}
$T_c$) would not obtain if fully renormalized Green's functions are
used everywhere\cite{Haussmann}, \cite{Tchern},\cite{Serene}.
Eqs.(\ref{Delta-def})-(\ref{number}) can be viewed as a simple (one
variable, $\Delta_{\rm pg}$) parameterization of Eqs. (\ref{Sigma}),
(\ref{T-matrix}), and (\ref{n}).

\begin{figure}
\narrowtext
\epsfxsize=3.3in 
\epsfbox{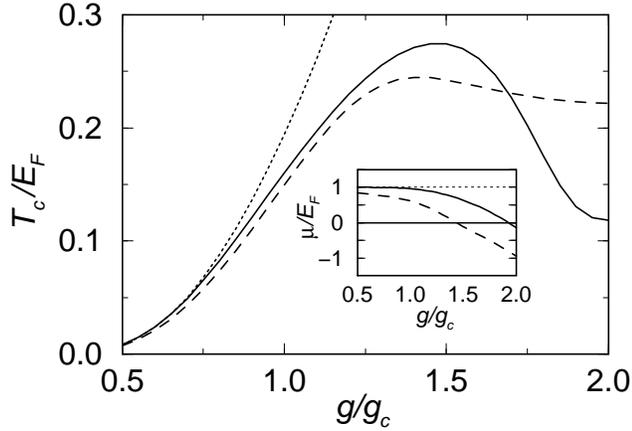}
\caption{Comparison of the $T_c$ curves within the BCS theory (dotted
line), the Nozi\`eres--Schmitt-Rink scheme (dashed line), and our mode
coupling results (heavy line). The inset shows the dependence of the
chemical potential $\mu (T=T_c,g)$ on the coupling constant for these
three cases.}
\label{fig:10}
\end{figure}

The overall behavior of $T_c$ is compared with that obtained from the
NSR approximation of Ref. ~\cite{NSR}, as well as that of strict BCS
theory in Figure ~\ref{fig:10}.  The non-monotonic behavior of $T_c$
results from the facts that (1) the high $g$ asymptote must be
approached from below\cite{Haussmann} and (2) as in Ref.\cite{NSR},
the low $g$ exponential dependence tends to overshoot this
asymptote. Here, the overshoot is even more marked than in the NSR
theory of Ref\cite{NSR} because of self energy effects which pin $E_F$
at $\mu$, as can be seen from the inset. Thus, at small $g$ our mode
coupling curves tend to follow the BCS result more closely.  Point (1)
is an important point. As $g$ increases, the pair size decreases and
consequently the Pauli principle repulsion is diminished.  In this way
$T_c$ must increase with $g$ at large $g$ in contrast to the behavior
found in Ref\cite{NSR}.

\begin{figure}
\narrowtext
\epsfxsize=3.3in 
\epsfbox{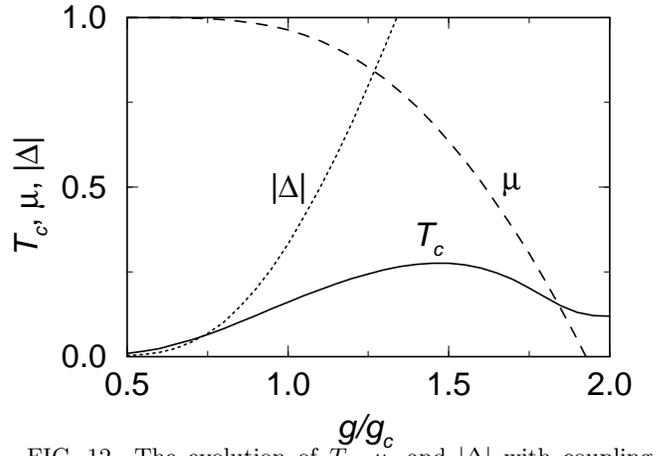}
\vspace{-0.1in}
\caption{The evolution of $T_c$, $\mu$, and $|\Delta|$ with
coupling strength in  self-consistent  mode coupling
theory.}
\label{fig:11}
\end{figure}

Figure \ref{fig:11} presents a comparison of the three different
energy scales: the chemical potential $\mu$, the pseudogap parameter $
\Delta_{\rm pg}$ and $T_c$.  It can be seen that the maximum in $T_c$
is associated with $\mu \sim \Delta_{\rm pg}$ and the minimum with
$\mu=0$.  Indeed, the complex behavior of $T_c$ shown in Figure
\ref{fig:10} can be understood on general physical grounds. A local
maximum appears in the $T_c$ curve as a consequence of a growing (with
increased coupling) pseudogap $\Delta_{\rm pg}$ in the fermionic
spectrum which weakens the superconductivity. However, even as
$\Delta_{\rm pg}$ grows, superconductivity is sustained.  In the
present scenario, superconductivity is preserved by the conversion of
an increasing fraction of fermions to bosonic states, which can then
Bose condense. Once the fermionic conversion is complete ($\mu=0$),
$T_c$ begins to increase again with coupling.

\begin{figure}
\narrowtext
\epsfxsize=3.3in 
\epsfbox{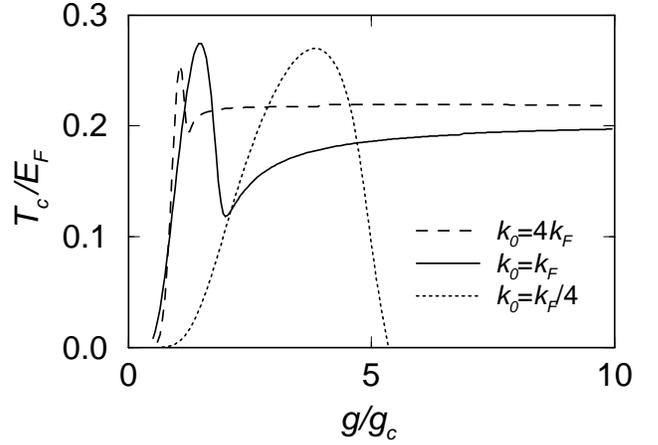}
\caption{ $T_c$ versus coupling constant over an extended range for 
three different values of the parameter $k_{\rm F}/k_0$. The 
minimum in the two larger $k_0$ curves is associated with  $ \mu = 
0$.}
\label{fig:12}
\end{figure}

The behavior of $T_c$ on an expanded coupling constant scale, for
different ranges of the interaction (parameterized by $k_0/k_{\rm F}$,
which enters via $\varphi_{\bf k} = (1+k^2/k^2_0)^{-1/2}$) is shown in
Figure \ref{fig:12}.  The limiting value of $T_c$ for large values of
$g/g_c$ approaches the ideal Bose-Einstein condensation temperature
$T_{\rm BE}=0.218E_{\rm F}$ for the case of short range attraction so
that $k_0\rightarrow\infty$.  The qualitative shape of the $T_c$
curve, however, is retained for $k_0/k_{\rm F}$ greater than about
0.5.  For larger range interactions, the solution disappears for some
$g/g_c$; then, when $\mu$ is sufficiently negative, the transition
reappears approaching a continuously (with ${k_0/k_{\rm
F}}\rightarrow0$) decreasing asymptote.

\subsection{Phase Diagram}

Our results for $T_c$ as a function of $g/g_c$ can be consolidated
into a quantitative phase diagram shown in Figure \ref{fig:13}, which
includes only the physical parameter regime ($\mu > 0 $). We, thus, do
not show larger values of the coupling constant where the system
acquires a bosonic or pre-formed pair character. The parameter $T^*$
is also indicated for each coupling constant.  This was determined as
the temperature at which a resonance at ${\bf q } = 0 $ first appears
in the T-matrix.  For lower values of the coupling constant $g / g_c <
1.0 $, this resonance onset essentially coincides with $T_c$, but
above this regime there is a clear separation between the two
temperature scales.  For $g/ g _c \approx 1.5 $ , as the pseudogap
size further increases, it starts to weaken the superconductivity, in
large part because of the decreased density of fermionic states which
can participate in the pairing.

There is, thus, a delicate balance between superconducting pairing and
the opening of a pseudogap associated with strong superconducting
attraction. In general gaps in the spectral function, as in charge or
spin density wave systems are expected to weaken
superconductivity. What is distinctive about a gap which arises from
the pairing, is that superconductivity persists even when the gap is
an order of magnitude greater than $T_c$.  This occurs in the present
model (as seen in Figure \ref{fig:11}) and this also occurs
experimentally\cite{arpesanl,arpesstanford} in the cuprates.
Superconductivity can coexist with large spectral function gaps when
these gaps are associated with a transition to a bosonic state: once
$\mu$ becomes negative with increasing $g$, the conversion of fermions
to bosons is complete and superconductivity, through Bose- Einstein
condensation, occurs in an unhindered fashion.  We view the
experimentally observed\cite{arpesanl,arpesstanford} large size of
$\Delta_{\rm pg} / T_c$ as suggestive of pseudogaps associated with
superconducting pairing. It is doubtful that other pseudogap
mechanisms would be as compatible with superconductivity.

Our phase diagram is plotted as a function $g/g_c$. The usual cuprate
phase diagram is plotted in terms of hole concentration dependence
$x$. This latter quantity has not been discussed thus far.
Nevertheless, there are two important effects which are associated
with the transition to the insulator $ x\rightarrow 0 $, which both
enhance pseudogap effects with decreasing $x$.  Here we assume for
definiteness that the pairing mechanism is not changed as the
insulator is approached. (This paper has not made any presumptions
about this mechanism and any associated hole concentration dependence
is not relevant to our considerations).  As the insulator is
approached, the system generally becomes progressively more two
dimensional, as seen for example, by the c-axis
resistivity\cite{CooperGray}.  The transition to two dimensionality is
associated\cite{Qinjin} with smaller and smaller $T_c$, so that the
size of the pseudogap temperature regime, which varies as $T^*/T_c$,
will grow as the insulator is approached.

Another important component of the transition to the insulator is the
decrease in plasma frequency. This issue has been emphasized by Emery
and Kivelson\cite{Emery}. Earlier work by our group\cite{MalyCoulomb}
demonstrated that when the Coulomb channel, along with charge effects,
were included into the Nozieres Schmitt-Rink formulation, they helped
to stabilize the mean field regime. The smaller the plasma frequency
the greater the deviation from BCS behavior.  This observation is not
readily associated with softening of phase fluctuations in the sense
of Ref. \cite{Emery}. In the present approach, the Coulomb channel
creates a particle-hole attraction which is competitive with the
particle particle (Cooper pair) attraction. In this way, one may infer
that as the insulator is approached, the system should become
progressively less mean field like, so that pseudogap effects are
expected to be enhanced. These and related effects will be discussed
in more detail in a companion paper\cite{Qinjin}.

\subsection{Relation to the literature}

In the previous section we discussed self energy effects, $T_c$
calculations and, in passing, some aspects of time dependent
Ginzburg--Landau (TDGL) theory which should be compared to related
calculations in the literature.

Our self energy calculations are in many ways similar to related work
on the effects on electronic properties of proximity to magnetic
transitions\cite{Brinkman,Vilk}. By contrast, in these other systems,
there have been no reports of a breakdown of the Fermi liquid phase,
except in the critical regime.  In most of these previous
calculations, as in some of ours, the relevant susceptibility was
expanded around the long wave length, low frequency limit. What causes
the Fermi liquid breakdown in our case, can be traced to a special
feature in this form of the pair susceptibility, which has no
counterpart in these other problems.  As noted earlier, such a
breakdown is naturally expected in any interpolation scheme which
varies smoothly between the fermionic and bosonic limits of $g/g_c$.
Our calculations show that the Fermi liquid like character in the self
energy is found to disappear as resonant structure in the T-matrix
sets in.  This can be seen directly from evaluating the integral
Eq. (\ref{ImSigma}) using Eq. (\ref{LG-exp}). Here resonance effects
enter via the ratio $(a'_0/a''_0)$. The larger is this parameter, the
more pronounced are pair resonances. Indeed, we are able to tune from
Fermi liquid to non Fermi liquid behavior simply by increasing the
size of this ratio, in very much the same way as was seen in the full
calculations, by increasing $g/g_c$.  For values of the ratio near
$\approx 1 $ the pseudogap state appears. Mode coupling effects which
renormalize $(a'_0/a''_0)$, can also be incorporated at this TDGL
level of approximation, as noted below.

Our TDGL expansion of the T-matrix, Eq. (\ref{LG-exp}) can be compared
to that discussed in the context of earlier work on the BCS
Bose-Einstein cross-over problem\cite{SadeMelo,Drechsler} It is well
known that as the Bose Einstein regime is approached, the parameter
$a'_0$ increases relative to $a''_0$. In this way the T-matrix
acquires a propagating component in addition to the diffusive term of
BCS theory. Indeed, this is precisely the situation in which the T
matrix is described by a resonance.  What is new in the present theory
is that, not only do we explicitly associate this resonance with the
break down of the Fermi liquid, but also for our mode coupling
calculations, the diffusive component (e.g., the parameter $a''_0$)
remains small for an extended range of temperatures above $T_c$ and
essentially vanishes at $T_c$ for all values of $g/g_c$. This
represents a restatement of our earlier conclusions that mode coupling
effects lead to extremely long lived pairs. These long lived pairs are
a reflection of the long lived single particle states whose rate of
decay is characterized by $\gamma$. Indeed, it can be shown by direct
expansion of the calculated T-matrix that
\begin{equation}
a''_0 \approx N(0)\frac{\gamma}{|\Delta_{\rm pg}|^2}.
\end{equation}
so that $a''_0$ varies as a power of $(T-T_c)$ sufficiently close to
$T_c$. This should be contrasted with the result
obtained\cite{SadeMelo,Drechsler} without mode coupling effects.

Finally, in the two dimensional (2d) limit our $T_c$
calculations\cite{Qinjin} can be compared with related work in the
literature. This 2d limit was noted to be problematic following
earlier work by Schmitt-Rink, Varma and Ruckenstein\cite{RVS}, who
applied the NSR theory to the low dimensional case. It was found that,
while $T_c$ was zero as expected, $\mu$ was also negative for all
values of $g/g_c$, including also in the weak coupling limit.
Serene\cite{Serene} pointed out this latter unphysical result arose
from the use of the truncated Dyson equation of Eq.  (\ref{DysonNSR}).
Yamada and co-workers \cite{Yamada} attempted to correct this problem
by including the lowest order ``box'' diagram in the T-matrix (See
Figure \ref{fig:0}(c)). In this way some mode coupling effects were
included; it was found that $\mu \approx E_F$ at arbitrarily weak
coupling, while $T_c$ was zero, as expected in the 2d limit. However,
the authors were unable to find a theory which smoothly interpolated
between the weak and strong coupling limits.

The present calculations (which, in contrast to earlier
``mode-coupling schemes'' of Ref. \cite{Marcelja,Schmid,Yamada},
include a series of higher order diagrams \cite{box}), yield
$T_c = 0 $, with $\mu$ varying continuously from $E_F$ to the large
negative values of the strong coupling limit. The vanishing of $T_c$
arises via Eq. (\ref{Delta-def}), which because of 2d phase space
restrictions on the Bose factor, yields an arbitrarily large
$\Delta_{\rm pg}$ which makes it impossible to satisfy the Thouless
criterion. In this way, problems with earlier 2d crossover theories are
corrected\cite{Qinjin}.

We end this discussion by noting that non monotonic behavior in $T_c$
with coupling $g$ has ample precedent in more conventional Fermi
liquid-based schemes, where it is found that, in some cases, $T_c$ may
decrease with increased large coupling constant, or saturate with
$g$. This follows within Eliashberg theory as a consequence of the
competing mass enhancement factor.  The resulting form for $T_c$
arises from a competition between the pairing coupling constant and
the mass enhancement ratio\cite{Levin}. The present calculations
report a similar effect, which arises, however, in the non Fermi
liquid regime. Here the competition is represented by the pseudogap
parameter $\Delta_{\rm pg}$ , which like $m^*/m$, can undermine the
effectiveness of the superconducting attraction.

\section{Conclusions}
\begin{figure}
\narrowtext
\epsfxsize=5.7in 
\epsfbox{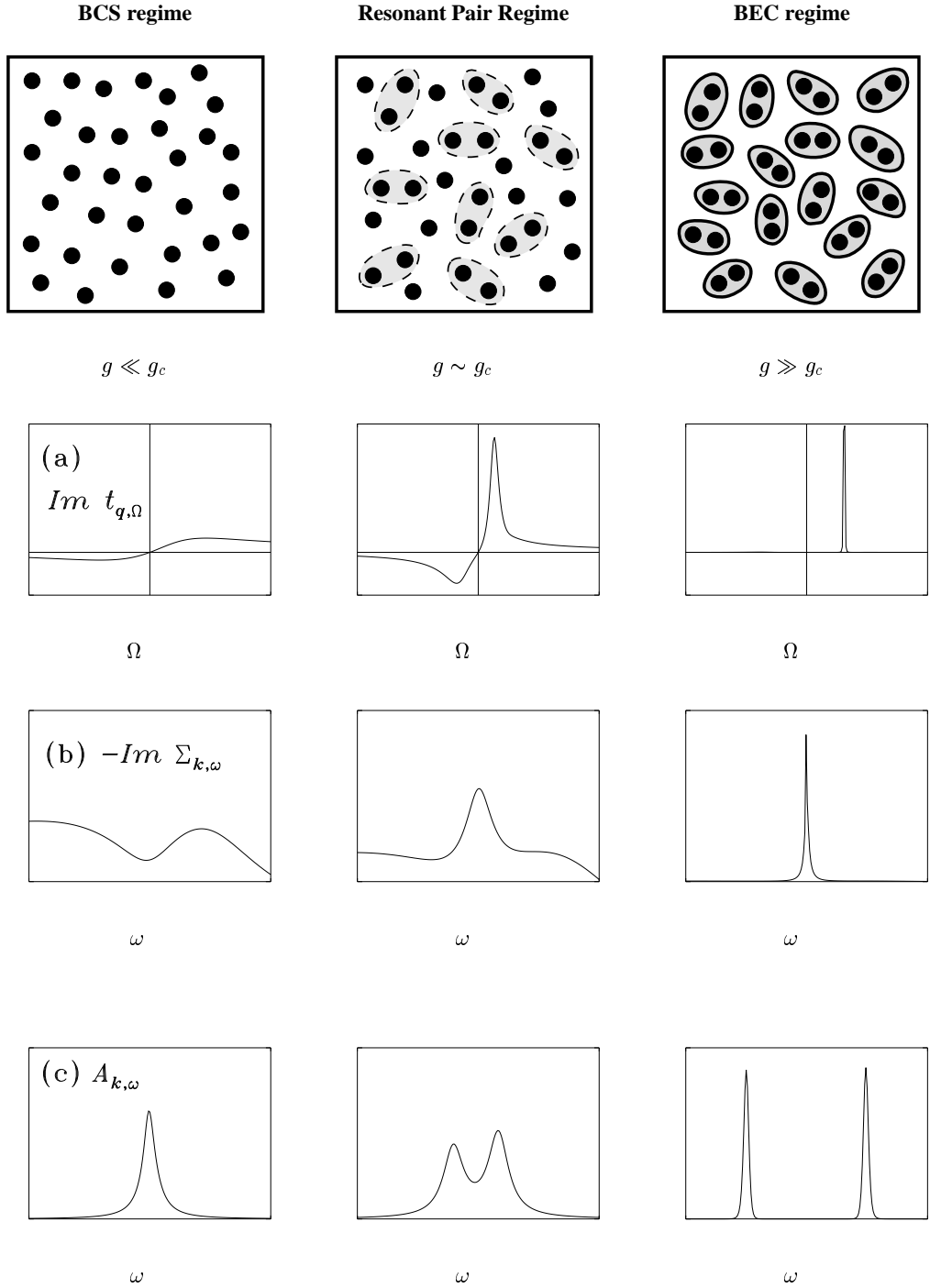}
\vspace{0.1in}
\caption{Summary of our conclusions: schematic plot of the resonant
pair or pseudogap regime as a crossover between free fermions (BCS)
and bound fermions (Bose-Einstein) with fermions indicated by the
solid dots, dashed lines around pairs denote metastable pair states,
while stable preformed pairs are represented by encircled pairs.
Panels (a), (b) and (c) illustrate the evolution with coupling of: (a)
${\it Im}\,t_{{\bf q=0},\Omega }$, (b) $-{\it Im}\Sigma_{{\bf
k},\omega}$, and (c) $A_{{\bf k},\omega}$. The vertical scales of
figures for BCS and resonance regime have appeared elsewhere,
throughout the paper. }
\label{fig:14}
\vspace{-0.1in}
\end{figure}

In this paper we have applied a diagrammatic decoupling scheme
proposed by Kadanoff and Martin, (which was further extended by
Patton), to the BCS- Bose Einstein cross over problem. The strongest
support for this theoretical framework is that it reproduces the BCS
limit at very weak coupling, in contrast to alternate diagrammatic
approaches in the literature.  The results of our extensive numerical
calculations are summarized in Figure \ref{fig:14}, in which we plot
our schematic physical picture, as well as the imaginary component of
the T-matrix, the imaginary part of the self energy and spectral
function for each of the three regimes: the weak coupling, BCS or
``Fermi liquid" case, the ``pseudogap state" and the ``bosonic
state'', or pre-formed pair limit. The physical picture of each
respective normal state corresponds to free fermions, resonantly
scattered fermions and bound fermions. The curves shown in the figure
for the first two cases represent quantitative plots, related to or
equivalent to those we have presented throughout this paper.

The present calculations have addressed a three dimensional $s$- wave
jellium model and have not made direct contact with the cuprate
superconductors. The effects of introducing $d$-wave pairing on a
quasi two dimensional lattice will be discussed
elsewhere\cite{Qinjin}: these effects do not qualitatively undermine
our conclusions.  Indeed, at this stage we can make preliminary
contact with experiments. In the pseudogap phase, our spectral
functions are generally consistent with a leading edge (albeit $s$-
wave) gap in the photoemission spectra. The gap, $\Delta_{\rm pg}$,
grows in magnitude with decreasing temperature, while at the same time
there is a narrowing of the associated spectral function peak, which
enters via the parameter $\gamma$. Precisely at $T_c$, $\gamma$ is
found to vanish, as appears to be the case
experimentally\cite{Norman}. Away from $T_c$ it grows rather slowly
with increasing temperature, presumably because of the assumption of
$s$- wave pairing.

\begin{figure}
\narrowtext
\epsfxsize=3.3in 
\epsfbox{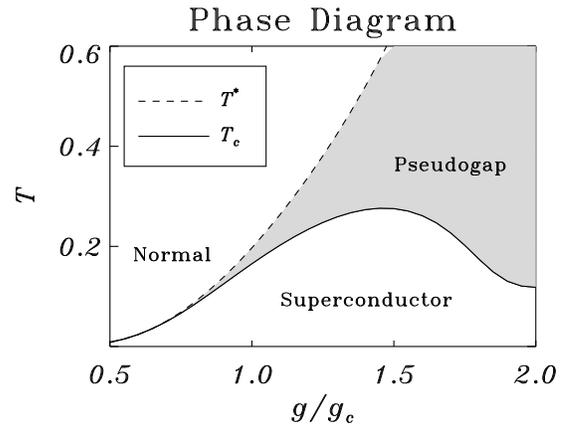}
\vspace{-0.1in}
\caption{Quantitative phase diagram in the  fermionic regime, $\mu 
>0$.  The shaded region indicates where there is a well established 
pseudogap in the single particle
spectral function.}
\label{fig:13}
\end{figure}

Our calculations have revealed a rich structure in the phase diagram
(See Figure \ref{fig:13}) for $T^*$, and $T_c$ as a function of
coupling constant. It is most likely that the cuprates are in the
regime of moderate coupling so that the pseudogap is not particularly
large compared to electronic energies \cite{arpesanl,arpesstanford}.
Combined with other experimental constraints, this appears consistent
with our model only in the quasi 2d limit\cite{Qinjin}.  However, it
must be stressed that, because it makes no assumption about the
pairing mechanism, the present theory is not appropriate for
addressing the detailed hole concentration dependence of the cuprate
phase diagram, except in this very qualitative fashion.  A more
detailed discussion of this and related issues is presented in a
companion paper\cite{Qinjin}.

Our diagrammatic calculations of the cross over from BCS to Bose
Einstein behavior, with increased coupling have several satisfying
features . Among these, we find (1) a well established pseudogap phase
with a spectral function which evolves continuously into the
characteristic two peaked structure of a conventional superconductor,
(2) that the Bose Einstein asymptote of our $T_c$ curves is properly
approached from below, (3) that a BCS limit is embedded in our weak
coupling expression for $T_c$, and (4) that, as will be discussed in
more detail in a subsequent paper, the 2d (and quasi 2d ) limits of
our theory appear to be well behaved, smoothly varying with coupling,
and yield a reasonable parameterization of cuprate energy
scales. Consequently, at the present time, we believe that these
cross-over theories, in the regime of intermediate coupling, should be
viewed as viable candidates for characterizing the precursor
superconductivity contributions to the cuprate pseudogap.

\acknowledgments

We would like to thank A. Abanov, V. Ambegaokar, A. S. Alexandrov,
B. L. Gyorffy, L. P. Kadanoff, A. Klein, P. A. Lee, P. B. Littlewood,
P. C. Martin, B. R. Patton, P. Wiegmann, A. Zawadowski and especially
to A. A. Abrikosov, Q.  Chen, I. Kosztin, M. Norman and Y. Vilk for
useful discussions.  This research was supported in part by the
Natural Sciences and Research Council of Canada (J.M.) and the Science
and Technology Center for Superconductivity funded by the National
Science Foundation under award No. DMR 91-20000.




\begin{references}
\bibitem{GGo} The detailed form used here for the T-matrix is more
readily related to that in Ref. \cite{Patton} than in
Ref.\cite{Kadanoff}.
\bibitem{wecheckedn} It should also be noted that we have verified
that this diagrammatic scheme is consistent with the general criteria
for number conservation in G. Baym and L. P. Kadanoff, Phys. Rev.
{\bf 124}, 287 (1961). Other conservation laws are also satisfied. For
the energy conservation see V. Ambegaokar and G. Rickayzen,
Phys. Rev. {\bf 142}, 146 (1966).
\bibitem{newfootnote}Here we have dropped a structureless term which
may be parameterized as $\Sigma _ {o} $. This is relatively
unimportant near $T_c$ and in the well developed pseudogap regime.  It
will play a role near $T^*$ when the cross-over to Fermi liquid
behavior occurs.  Moreover, its presence is necessary to "regularize"
the self energy when $\gamma$ is arbitrarily small so that self energy
is never a strict delta function in the normal state.
\end{references}

\vspace*{-1ex}

\vspace{-0.2in}

\vspace{-0.2in}

\vspace{-0.2in}

\vspace{-0.3in}

\end{document}